\documentclass[12pt,graphics]{iopart}

\usepackage{iopams}  
\usepackage{graphicx}
\usepackage{marvosym}        
\usepackage{epsfig}
\usepackage{units}
\usepackage{multicol,graphics,verbatim}
\usepackage{longtable}
\usepackage{colortbl}
\usepackage{color}
\usepackage{rotating}  
\usepackage{nicefrac} % for nice 1/2 sign
\usepackage{calc} % for parbox = including indented paragraph
\usepackage{amssymb}
\usepackage{amsthm}
\usepackage{upgreek}          %  for upright greek letters in units
\usepackage{ifthen}
\usepackage[colorinlistoftodos]{todonotes}
\usepackage{threeparttable}
\newcommand{\bpbp}{\mbox{$\beta^+\beta^+$} }
\newcommand{\ecec}{\mbox{$\rm ECEC$} }
\newcommand{\bec}{\mbox{$\rm \beta^+EC$} }
\newcommand{\bnel}{\mbox{$\bar{\nu}_e$} }

\newcommand{\nel}{\mbox{$\nu_e$}}

\newcommand{\be}{\begin{equation}}
\newcommand{\ee}{\end{equation}}
\def\bea{\begin{eqnarray}} 
\def\eea{\end{eqnarray}} 
\newcommand{\ra}{\rightarrow }

\begin{document}

\newcommand{\nuc}[2]{$^{#2}\rm #1$}

\newcommand{\bb}[1]{$\rm #1\nu \beta \beta$}
\newcommand{\bbm}[1]{$\rm #1\nu \beta^- \beta^-$}
\newcommand{\bbp}[1]{$\rm #1\nu \beta^+ \beta^+$}
\newcommand{\bbe}[1]{$\rm #1\nu \rm ECEC$}
\newcommand{\bbep}[1]{$\rm #1\nu \rm EC \beta^+$}

\newcommand{\gerda}{\textsc{Gerda}}
\newcommand{\largeGERDA}{{LArGe}}
\newcommand{\PI}{\mbox{\textsc{Phase\,I}}}
\newcommand{\PIa}{\mbox{\textsc{Phase\,I}a}}
\newcommand{\PIb}{\mbox{\textsc{Phase\,I}b}}
\newcommand{\PIc}{\mbox{\textsc{Phase\,I}c}}
\newcommand{\PII}{\mbox{\textsc{Phase\,II}}}

\newcommand{\geant}{\textsc{GEANT4}}
\newcommand{\mage}{\textsc{MaGe}}

\newcommand{\nPlus}{\mbox{n$^+$ electrode}}
\newcommand{\pPlus}{\mbox{p$^+$ electrode}}

\newcommand{\AOE}{$A/E$}

\newcommand{\order}[1]{\mbox{$\mathcal{O}$(#1)}}

\newcommand{\mul}[1]{\texttt{multiplicity==#1}}

\newcommand{\baseT}[2]{\mbox{$#1\cdot10^{#2}$}}
\newcommand{\baseTsolo}[1]{$10^{#1}$}
\newcommand{\THL}{$T_{\nicefrac{1}{2}}$}

\newcommand{\UBI}{$\rm cts/(kg \cdot yr \cdot keV)$}

\newcommand{\Uflux}{$\rm m^{-2} s^{-1}$}
\newcommand{\Ucpd}{$\rm cts/(kg \cdot d)$}
\newcommand{\Uexpo}{$\rm kg \cdot d$}
\newcommand{\UexpoYear}{$\rm kg \cdot yr$}

\newcommand{\UMWE}{m.w.e.}

\newcommand{\Qbb}{$Q_{\beta\beta}$}

\newcommand{\validate}{\textcolor{blue}{\textit{(validate!!!)}}}

\newcommand{\improve}{\textcolor{blue}{\textit{(improve!!!)}}}

\newcommand{\missing}{\textcolor{red}{\textbf{...!!!...} }}

\newcommand{\quanta}{\textcolor{red}{\textit{(quantitativ?) }}}

\newcommand{\misscite}{\textcolor{red}{[citation!!!]}}

\newcommand{\missref}{\textcolor{red}{[reference!!!]}\ }

%K42
\newcommand{\PC}{$N_{\rm peak}$}
\newcommand{\BIC}{$N_{\rm BI}$}
\newcommand{\PAPR}{$R_{\rm p/>p}$}

\newcommand{\PCR}{$R_{\rm peak}$}

%Pd

\newcommand{\gline}{$\gamma$-line}
\newcommand{\glines}{$\gamma$-lines}

\newcommand{\gray}{$\gamma$-ray}
\newcommand{\grays}{$\gamma$-rays}

\newcommand{\bray}{$\beta$-ray}
\newcommand{\brays}{$\beta$-rays}

\newcommand{\aray}{$\alpha$-ray}
\newcommand{\arays}{$\alpha$-rays}

\newcommand{\betas}{$\beta$'s}

%general

\newcommand{\tab}{{Tab.~}}
\newcommand{\eq}{{Eq.~}}
\newcommand{\fig}{{Fig.~}}
\renewcommand{\sec}{{Sec.~}}
\newcommand{\chap}{{Chap.~}}

 \newcommand{\fn}{\iffalse \fi} %footnote explaination
 \newcommand{\tx}{\iffalse \fi} %text explaination
 \newcommand{\txe}{\iffalse \fi} %text extended explaination
 \newcommand{\sr}{\iffalse \fi} %section reference explaination

\today

% * <thomas.wester@tu-dresden.de> 2015-08-28T12:11:02.721Z:
%
% 
%
\title{Double Beta Decays into Excited States in $^{110}$Pd and $^{102}$Pd}
%Possibility of measuring long-living \zbb emitters in the presence of short-living ones or }

\author{B. Lehnert$^a$, E. Andreotti$^b$, D. Degering$^c$, M. Hult$^b$,\\ M. Laubenstein$^d$, T. Wester$^a$, K. Zuber$^a$}
\address{
$^a$ Institut f\"ur Kern- und Teilchenphysik, Technische Universit\"at Dresden, Zellescher Weg 19, 01069 Dresden, Germany,\\
$^b$ Institute for Reference Materials and Measurements, Retieseweg 111, B-2440 Geel, Belgium,\\
$^c$ VKTA - Radiation Protection, Analytics \& Disposal Rossendorf e.V., P.O.Box 510119, 01314 Dresden, Germany\\
$^d$ INFN - Laboratori Nazionali del Gran Sasso, S.S. 17 bis km 18+910, Assergi (AQ), Italy\\
}
\ead{bjoernlehnert@gmail.com}

\begin{abstract}
A search for double beta decays of \nuc{Pd}{110} and \nuc{Pd}{102} into excited states of the daughter nuclides  has been performed using three ultra-low background gamma-spectrometry measurements in the Felsenkeller laboratory, Germany, the HADES laboratory, Belgium and at the LNGS, Italy. The combined Bayesian analysis of the three measurements sets improved half-life limits for the \bb{2} and \bb{0} decay modes of the $2^+_1$, $0^+_1$ and $2^+_2$ transitions in \nuc{Pd}{110} to 
\unit[\baseT{2.9}{20}]{yr}, \unit[\baseT{4.0}{20}]{yr} and \unit[\baseT{3.0}{20}]{yr} respectively and in \nuc{Pd}{102} to \unit[\baseT{7.6}{18}]{yr}, \unit[\baseT{8.8}{18}]{yr} and \unit[\baseT{1.4}{19}]{yr} respectively with \unit[90]{\%} credibility.
\end{abstract}

\maketitle

\section{Introduction}
%\linenumbers
%\pagewiselinenumbers

The investigation of neutrinoless double beta (\bb{0}) decay is a promising approach to search for physics beyond the Standard Model (SM). This second order weak nuclear decay requires lepton number violation which can in principle be generated by many new theories. As a consequence also the Majorana nature of the neutrino would be implied. The most intuitive mechanism to describe the decay is the exchange of a virtual light Majorana neutrino linking its mass ($m_{ee}$) to the half-life ($T_{1/2}^{0\nu}$) of the decay: 
\begin{eqnarray}
 0\nu\beta\beta: \left(T_{1/2}^{0\nu}\right)^{-1} = G^{0\nu} \cdot \left|M^{0\nu}\right|^2 \cdot \left|m_{ee}\right|^2\, .\label{eq:0nbb}
\end{eqnarray}

A phase space factor (PSF, $G^{0\nu}$) and a nuclear matrix element (NME, $M^{0\nu}$) are required for the conversion and both are strongly depending on the nuclide under study. While the calculation of the PSF is relatively straight forward \cite{Kotila:2012bj}, the calculation of the NME is subject to large theoretical uncertainties. Various nuclear models are applied for the calculations which currently disagree by around a factor of three \cite{Barea:2015jz} and are the largest uncertainty in constraining the effective Majorana neutrino mass with double beta decay experiments \cite{Suhonen:2012dn}. 
The nuclear model calculations might be improved by providing additional experimental information in the same nuclear systems. This is possible with the investigation of the SM allowed process of two neutrino double beta (\bb{2}) decay which has been well observed in over 10 nuclides. In contrast to the \bb{0} mode, the partial half-life for the \bb{2} mode can be calculated directly via a PSF and NME:
\begin{eqnarray}
2\nu\beta\beta: \left(T_{1/2}^{2\nu}\right)^{-1} &= G^{2\nu} \cdot \left|M^{2\nu}\right|^2\, . \label{eq:2nbb}
\end{eqnarray}

Although the calculation of NMEs for the \bb{2} and \bb{0} modes are based on different nuclear levels in the intermediate nucleus and are numerically different, they can nevertheless be obtained in the same nuclear model framework. Experimental data on \bb{2} decay helps to verify the calculation of \eq (\ref{eq:2nbb}) in a given framework and thus creating confidence in the calculations for \eq (\ref{eq:0nbb}). In addition, free nuclear model parameters can be constrained \cite{Barea:2015jz}. For recent reviews see \cite{AvignoneIII:2008wm}.\\

Apart from decays into the ground state (g.s.), double beta decays can also occur into the excited states of the daughter nucleus. These decay modes are expected to have a slower rate due to a smaller phase space but their experimental signature is enhanced by accompanying de-excitation \grays. Excited state transitions can in principle occur in the \bb{2} and the \bb{0} domain with difference only in the residual electron energy. The investigation of \bb{2} modes into excited states provides additional information on the nuclear structure and can over-constrain the system \eq (\ref{eq:0nbb},\ref{eq:2nbb}). So far only transitions to the first excited $0^+_1$ state have been observed, in \nuc{Mo}{100} \cite{Barabash:1995jt,Barabash:1999dp,Arnold:2006fk,Kidd:2009ai,Belli:2010zzc} and in \nuc{Nd}{150} \cite{Barabash:2004vf} with recent half-life values of \THL\unit[\baseT{=(7.5\pm1.2)}{20}]{yr} \cite{Nemo3:2014iy} and \THL\unit[$=\baseT{(1.33^{+0.63}_{-0.36})}{20}$]{yr} \cite{Barabash:2009fw}, respectively. The half-life calculations of the ground state and excited state transitions in these nuclides based on the same nuclear model parameters are currently inconsistent by more than one order of magnitude \cite{Lehnert:2015el}.\\

The equivalent processes to double beta minus decay with the emission of two electrons could also occur on the right side of the mass parabola of even-even isobars. Three different decay modes are possible involving $\beta^+$ decay and electron capture (EC)
\begin{eqnarray}
\label{eq:ecec}
\mbox{$0\nu[2\nu]$$\beta^-\beta^-$}&:\quad  (Z,A) &\ra (Z+2,A) + 2 e^-  [+ 2 \bnel]\\[1pc]
\mbox{$0\nu[2\nu]$\ecec}&:\quad  2 e^- + (Z,A) &\ra (Z-2,A) \; [+ 2 \nel]  \label{eqn:ecb+}\\
\mbox{$0\nu[2\nu]$\bec}&:\quad   e^- + (Z,A) &\ra (Z-2,A) + e^+ \; [+ 2 \nel]   \label{eqn:ecec}\\
\mbox{$0\nu[2\nu]$\bpbp}&:\quad    (Z,A) &\ra  (Z-2,A) + 2 e^+ \; [+ 2 \nel] \label{eqn:b+b+} 
\end{eqnarray}
Decay modes containing an EC emit X-rays or Auger electrons created by the atomic shell vacancy in the daughter nuclide. Decay modes containing a $\beta^+$ create two \unit[511]{keV} annihilation \grays\ and have a reduced phase space by \unit[1022]{keV} per $\beta^+$.\\

%
%
%\be
%(Z,A) \ra (Z+2,A) + 2 e^-  \quad (\obb).
%\ee   
%
%
%\be
%(Z,A) \ra (Z+2,A) + 2 e^-  + 2 \bnel \quad (\zbb)
%\ee      
%
%
%\be
%\left(\tzn\right)^{-1} = G \times \mid M_{GT} \mid^2\,,
%\ee
%
%
%\begin{eqnarray}
%\label{eq:ecec}
%(Z,A) &\ra  (Z-2,A) + 2 e^+ \; (+ 2 \nel) \quad & \mbox{(\bpbp)} \label{eqn:b+b+}\\
%e^- + (Z,A) &\ra (Z-2,A) + e^+ \; (+ 2 \nel) \quad & \mbox{(\bec)} \label{eqn:ecec}\\
%2 e^- + (Z,A) &\ra (Z-2,A) \; (+ 2 \nel) \quad & \mbox{(\ecec)} \label{eqn:ecb+}
%\end{eqnarray}
%
%

\begin{figure}
\centering
\includegraphics[width=0.9\columnwidth]{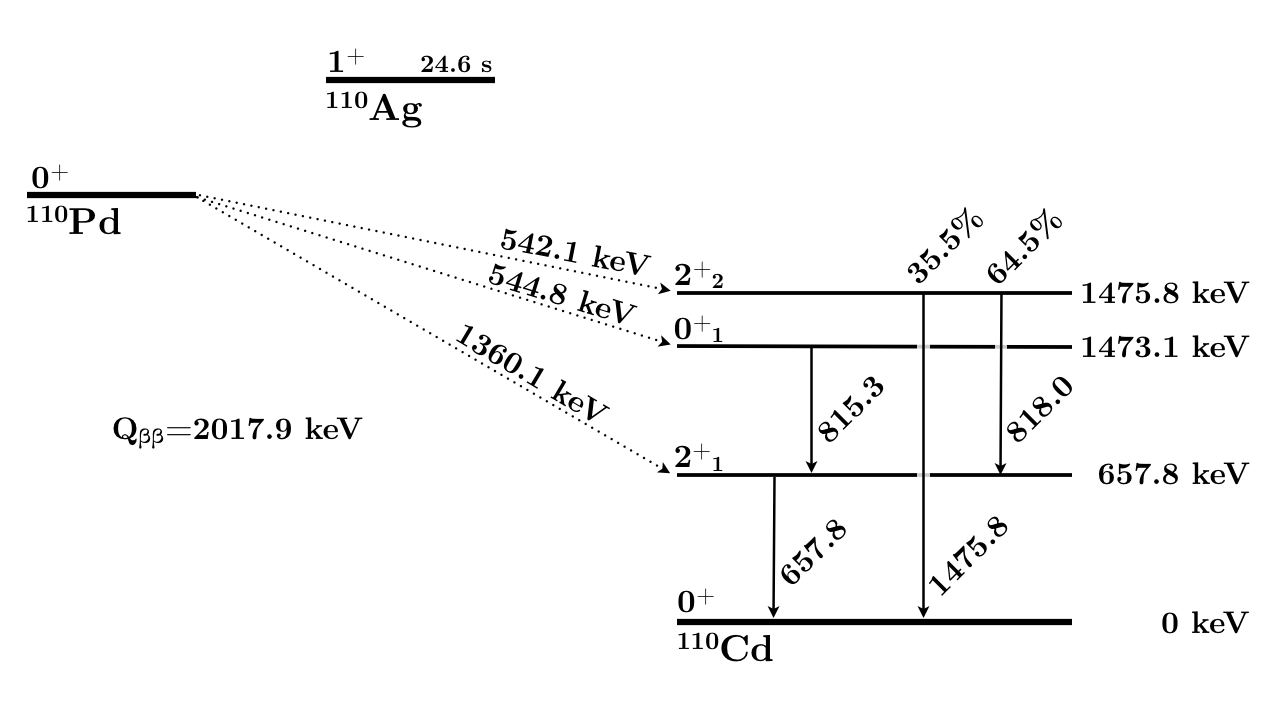}
\caption{Decay scheme of \nuc{Pd}{110}. Nuclear data from \cite{website:NuDat2}.}
\label{fig:decayschemePd110}
\end{figure}

\begin{figure}
\centering
\includegraphics[width=0.9\columnwidth]{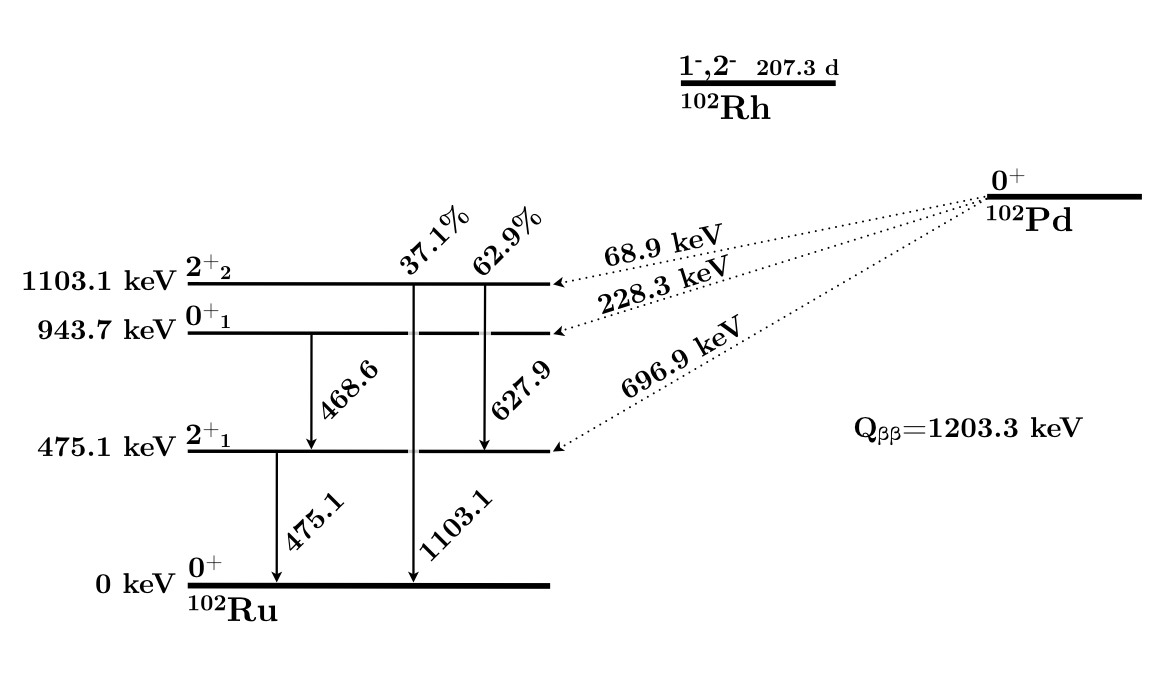}
\caption{Decay scheme of \nuc{Pd}{102}. Nuclear data from \cite{website:NuDat2}.}
\label{fig:decayschemePd102}
\end{figure}

The element under study is palladium with the isotopes of interest \nuc{Pd}{110} and \nuc{Pd}{102}. The decay schemes including the investigated transitions are shown in \fig \ref{fig:decayschemePd110} and \fig \ref{fig:decayschemePd102}, respectively. Among the 35 nuclides expected to undergo $\beta^-\beta^-$ decay, \nuc{Pd}{110} has the second highest natural abundance with \unit[11.72]{\%}. Recently, the Q-value was remeasured to \unit[2017.85(64)]{keV} \cite{Fink:2012hz} and places \nuc{Pd}{110} among the 11 $\beta^-\beta^-$ nuclides with a Q-value larger than \unit[2]{MeV}. The second isotope \nuc{Pd}{102} has a Q-value of \unit[1203.27(36)]{keV} \cite{Goncharov:2011dhb}, a natural abundance of \unit[1.02]{\%} and is able to decay via $0\nu[2\nu]$\ecec and $0\nu[2\nu]\beta^+$EC.\\

Three measurements of \nuc{Pd}{110} have been performed in the past, one in 1954 \cite{Winter:1952fe} and two more recently in 2011 in the Felsenkeller laboratory \cite{Lehnert:2011bw} and in 2013 in the HADES laboratory \cite{Lehnert:2013ch}. The latter two measurements were the first to investigate excited state transitions in palladium with gamma-spectrometry. The search described here is based on a combination of data from these previous measurements and data from a new measurement at the Laboratory Nazionali del Gran Sasso (LNGS).
For \bb{2} \nuc{Pd}{110} excited state transitions exist many theoretical calculations to which the experimental limits can be compared. For the \bb{2} \nuc{Pd}{102} excited state transitions only experimental half-life limits are known and no theoretical calculation have been published up to date. 
The existing experimental and theoretical half-life limits are summarized in \tab \ref{tab:table2}.

\begin{table}[h]
\caption{\label{tab:table2} Theoretical half-life predictions and experimental limits for \bbm{2} decays of \nuc{Pd}{110} and \bbe{2} decays of \nuc{Pd}{102} into various excited state modes. The experimental limits are also valid for the \bb{0} decay modes. The columns show from left to right the quoted half-life, the theoretical model, the reference and the year of publication. Abbreviations denote:
%PHFM - Projected Hartree-Fock-Bogoliubov,
SRPA - secon random phase approximation,
SSD - single state dominance,
%OEM - Operator Expansion Method,
%QRPA - Quasi Random Phase Approximation and
pnQRPA - proton-neutron quasiparticle random-phase approximation and
IBM-2 - interacting boson model.
}
\begin{center}
\footnotesize
\begin{tabular}{l|cccc}
\br
\bb{2} decay & \THL\ [yr]& model & ref. & year\\
\mr
%\nuc{Pd}{110} $0_{\rm g.s.}^+ - 0^+_{\rm g.s.}$
% & \baseT{1}{17} (\unit[68]{\%} CL) &  		exp. & \cite{Winter:1952fe}  & 1952\\
% &  \baseT{1.16}{19}   & QRPA &  \cite{Staudt:1990jw} & 1990\\
% &  \baseT{1.6}{20}    &  SRPA & \cite{Stoica:1994gz} & 1994\\
% &  \baseT{1.24}{21}    & OEM &  \cite{Hirsch:1994kz} & 1994 \\
% &  \baseT{1.2-1.8}{20} $^a$    & SSD &  \cite{Civitarese:1998hn} & 1998 \\
% &  \baseT{1.75}{20}  &  SSD & \cite{Semenov:2000sv}  & 2000\\
% &   \baseT{1.2}{20}   & SSD &  \cite{Domin:2005dl} & 2005\\
% &  \baseT{1.41}{20} and \baseT{3.44}{20} $^b$  & PHFM &  \cite{Chandra:2005gb} & 2005\\
% &   \baseT{1.1}{20} and \baseT{0.91}{20} $^c$ & pnQRPA  & \cite{Suhonen:2011du} & 2011 \\
% &  \baseT{1.5}{20} $^d$ & IBM-2 &  \cite{Barea:2015jz} & 2015 \\
%\mr
\nuc{Pd}{110} $0_{\rm g.s.}^+ - 2^+_1$ (\unit[657.76]{keV})
 &  \baseT{4.40}{19} (\unit[95]{\%} CL)  &  exp. & \cite{Lehnert:2011bw} & 2011\\
 &   \baseT{1.72}{20} (\unit[95]{\%} CL)   &  exp. & \cite{Lehnert:2013ch}  & 2013 \\
 &   \baseT{8.37}{25}  & SRPA  & \cite{Stoica:1994gz} & 1994 \\
 &  \baseT{4.4}{25}   &  SSD & \cite{Domin:2005dl}  & 2005\\
 &   \baseT{1.48}{25} & pnQRPA  &  \cite{Raduta:2007hwa}  & 2007\\
 &   \baseT{0.62}{25} and \baseT{1.3}{25} $^a$  & pnQRPA   &  \cite{Suhonen:2011du} & 2011 \\
\mr
\nuc{Pd}{110} $0_{\rm g.s.}^+ - 0^+_1$ (\unit[1473.12]{keV})
 &  \baseT{5.89}{19} (\unit[95]{\%} CL) & exp. &  \cite{Lehnert:2011bw} & 2011\\
 &   \baseT{1.98}{20} (\unit[95]{\%} CL)  &  exp. & \cite{Lehnert:2013ch} & 2013 \\
 &  \baseT{2.4}{26}  & SSD  &  \cite{Domin:2005dl} & 2005\\
 &  \baseT{4.2}{23} and \baseT{9.1}{23} $^a$  & pnQRPA &  \cite{Suhonen:2011du} &2011 \\
 &  \baseT{2.9}{26} $^b$ & IBM-2 &  \cite{Barea:2015jz} & 2015 \\
\mr
\nuc{Pd}{110} $0_{\rm g.s.}^+ - 2^+_2$ (\unit[1475.80]{keV})
 &   \baseT{9.26}{19} (\unit[95]{\%} CL)  & exp. &  \cite{Lehnert:2013ch}  & 2013\\
 &  \baseT{3.8}{31}  & SSD &  \cite{Domin:2005dl} & 2005\\
 &   \baseT{11}{30} and \baseT{7.4}{30} $^a$ & pnQRPA &  \cite{Suhonen:2011du} & 2011\\
\mr
\nuc{Pd}{110} $0_{\rm g.s.}^+ - 0^+_2$ (\unit[1731.33]{keV})
 &   \baseT{1.38}{20} (\unit[95]{\%} CL) &  exp. &  \cite{Lehnert:2013ch}  & 2013\\
 &   \baseT{5.3}{29}  & SSD &   \cite{Domin:2005dl} & 2005\\
\mr
\nuc{Pd}{110} $0_{\rm g.s.}^+ - 2^+_3$ (\unit[1783.48]{keV})
 &   \baseT{1.09}{20} (\unit[95]{\%} CL) & exp. &  \cite{Lehnert:2013ch} & 2013 \\
 &  \baseT{1.3}{35}  &  SSD & \cite{Domin:2005dl}  & 2005\\
\mr
\mr
\nuc{Pd}{102} $0_{\rm g.s.}^+ - 2^+_{1}$ (\unit[475.10]{keV})
 &   \baseT{2.68}{18} (\unit[95]{\%} CL)  & exp. & \cite{Lehnert:2011bw} &2011 \\
 &   \baseT{5.95}{18} (\unit[95]{\%} CL)  & exp. & \cite{Lehnert:2013ch}  & 2013\\
\mr
\nuc{Pd}{102} $0_{\rm g.s.}^+ - 0^+_{1}$ (\unit[943.69]{keV})
 & \baseT{7.64}{18} (\unit[95]{\%} CL)   & exp. & \cite{Lehnert:2011bw}  &2011\\
 &  \baseT{5.81}{18} (\unit[95]{\%} CL)   & exp. & \cite{Lehnert:2013ch}  & 2013\\
\mr
\nuc{Pd}{102} $0_{\rm g.s.}^+ - 2^+_{2}$ (\unit[1103.05]{keV})
 &   \baseT{8.55}{18} (\unit[95]{\%} CL)   & exp. & \cite{Lehnert:2013ch}  &2013\\
\br 
\end{tabular}\\
\end{center}
\footnotesize
$^a$ {For Woods-Saxon Potential and adjusted base respectively. See \cite{Suhonen:2011du} for details}\\
$^b$ {Calculation based on PSF from \cite{Kotila:2012bj} and NME from \cite{Barea:2015jz}. See \cite{LehnertPhD} for details. }

\end{table}

%%%%%%%%%%%%%%%%%%%%%%%%%%%%%%%%%%%%%%%%%%%%%%%%%%%%%%%%%%%%%%%%%%%%%%%%%%%%%%%%%%%%%%%%%%%%%%%%%%%%%%%
%%%%%%%%%%%%%%%%%%%%%%%%%%%%%%%%%%%%%%%%%%%%%%%%%%%%%%%%%%%%%%%%%%%%%%%%%%%%%%%%%%%%%%%%%%%%%%%%%%%%%%%
%%%%%%%%%%%%%%%%%%%%%%%%%%%%%%%%%%%%%%%%%%%%%%%%%%%%%%%%%%%%%%%%%%%%%%%%%%%%%%%%%%%%%%%%%%%%%%%%%%%%%%%

\section{Palladium Sample}
\label{sample}
%# small std boxes OD 55 mm, id 53 mm, h 3.0 mm, bottom and top 1.0 mm; 
%# lower box is full: (486.8 - 9.2) g = 477.6 g, upper box filled with two layers: (344.1 - 9.0) g = 335.1 g 
%# start of measurement 21-AUG-2012

The sample consists of \unit[802.4]{g} of irregularly shaped \unit[$1$]{mm} x \unit[1]{cm$^2$} palladium plates. 
Prior to any of the measurements, the sample was purified by C. HAFNER GmbH + Co. KG in 2010 to a certified purity of \unit[$>99.95$]{\%} which lowered the continuous background in the peak regions by approximately \unit[20]{\%} \cite{Lehnert:2011bw}. In order to avoid radionuclides produced by cosmic ray spallation, the palladium was kept underground apart from \unit[$3$]{weeks} during purification in 2010, \unit[$3$]{weeks} for surface transport in fall 2011, and \unit[$2$]{days} transport in spring 2012 of which \unit[3]{hours} were done by airplane. \\

\begin{figure}[h]
\centering
                \includegraphics[width=\textwidth]{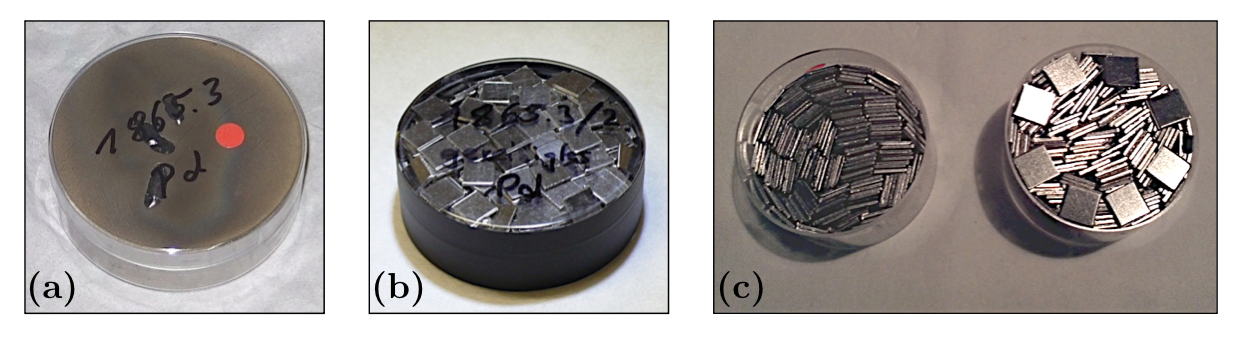}
         \caption{\label{fig:PdSample} Palladium sample prior to purification (a), for the Felsenkeller and HADES measurements (b) and for the LNGS measurement (c).}
\end{figure}

For the measurements at the Felsenkeller and in HADES the palladium was placed in a single container of polystyrene with \unit[70]{mm} diameter and \unit[21]{mm} height. The effective density is calculated as \unit[10.2]{g/cm$^3$} compared to the bulk density of palladium of \unit[12.02]{g/cm$^3$}.
For the measurement at LNGS the plates were placed inside two measuring containers of \unit[55]{mm} diameter and \unit[30]{mm} height which were piled on top of each other. The effective density is calculated as \unit[7.59]{g/cm$^3$}. A picture of the palladium sample is shown in \fig \ref{fig:PdSample} before purification (a), as used for the Felsenkeller and HADES measurements (b) and as used for the LNGS measurement (c).\\

The radioactive impurity of the palladium sample was assessed during the gamma-spectrometry measurement in HADES and positive evidence for \nuc{Pb}{214} and \nuc{Bi}{214} was found with around \unit[2]{mBq/kg} activity \cite{Lehnert:2013ch}. This indicates the presence of \nuc{Ra}{226} in either the sample (most likely) or the container. Additionally, a potential contamination of the following radionuclides was investigated: \nuc{Rh}{102} (\THL = \unit[207.3]{d}), \nuc{Rh}{\rm 102m} (\THL = \unit[3.742]{yr}) and \nuc{Ag}{\rm 110m} (\THL = \unit[249.76]{d}). The reason is the possible interference with the search for \nuc{Pd}{110} and \nuc{Pd}{102} decays, because of the emission of \grays\ from the same excited daughter states. The investigation was performed with additional \glines\ of these decays which are not part of the experimental signal of the double beta decay transitions due to the larger Q-value of the beta decay and the additional EC. No presence of these radionuclides was found.

%%%%%%%%%%%%%%%%%%%%%%%%%%%%%%%%%%%%%%%%%%%%%%%%%%%%%%%%%%%%%%%%%%%%%%%%%%%%%%%%%%%%%%%%%%%%%%%%%%%%%%%
%%%%%%%%%%%%%%%%%%%%%%%%%%%%%%%%%%%%%%%%%%%%%%%%%%%%%%%%%%%%%%%%%%%%%%%%%%%%%%%%%%%%%%%%%%%%%%%%%%%%%%%
%%%%%%%%%%%%%%%%%%%%%%%%%%%%%%%%%%%%%%%%%%%%%%%%%%%%%%%%%%%%%%%%%%%%%%%%%%%%%%%%%%%%%%%%%%%%%%%%%%%%%%%
\section{Experimental Setup and Datasets}
\label{setup}

\begin{figure}[ht]
\centering
                \includegraphics[width=0.8\textwidth]{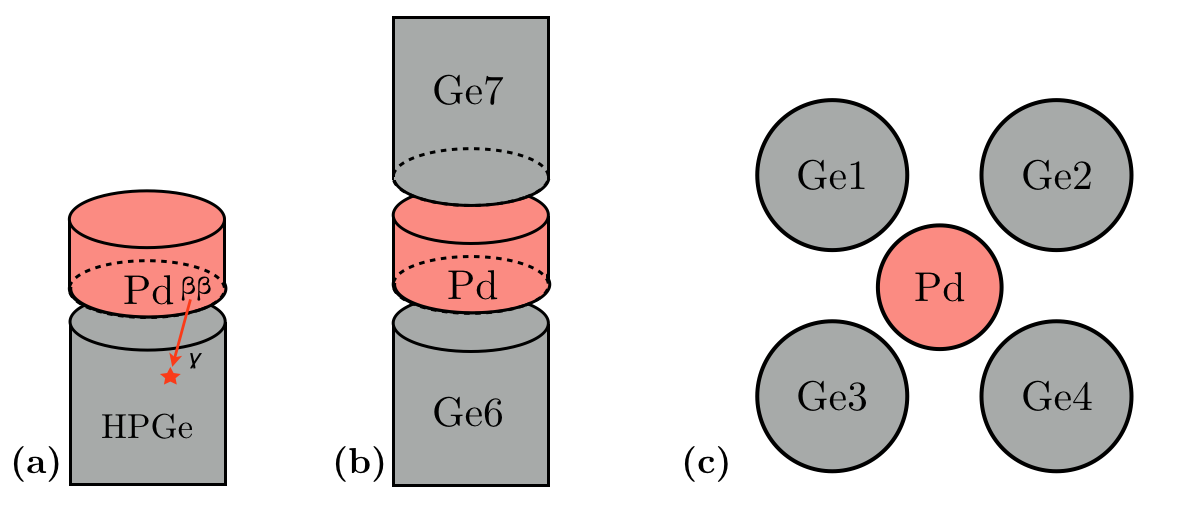}
         \caption{\label{pic:PdCombined_detScheme} Detector and sample configuration for the Felsenkeller (a), HADES (b) and LNGS (c) setup. The two electrons of the \nuc{Pd}{110} decay remain in the sample and the \bb{2} and \bb{0} decay mode cannot be distinguished.}
\end{figure}

The data used in this work comprises of three datasets obtained with three different ultra-low background gamma-spectrometry setups which are illustrated in \fig \ref{pic:PdCombined_detScheme}. 

%The details of the Felsenkeller and HADES setups and measurements can be found in \cite{Lehnert:2011bw} and \cite{Lehnert:2013ch} respectively and are briefly summarized below.

%%%%%%%%%%%%%%%%%%%%%%%%%%%%%%%%%%%%%%%%%%%%%%%%%%%%%%%%%%%%%%%%%%%%%%%%%%%%%%%%%%%%%%%%%%%%%%%%%%%%%%%
\subsection{Felsenkeller Dataset}

The measurement was performed in the Felsenkeller underground laboratory in Dresden, Germany, with a shielding of \unit[110]{m.w.e.}\ rock overburden reducing the muon flux by about a factor of 20 to \unit[${0.6}\cdot 10^{-3}$]{cm$^{-2}$s$^{-1}$} \cite{Niese:1998va,FelixDPG}. The palladium sample was measured for 16.2 days (\unit[13.0]{kg$\cdot$d} exposure) with a \unit[90]{\%} efficiency HPGe detector routinely used for gamma-spectrometry \cite{Kohler:2009ena}.
The  detector is surrounded by a \unit[5]{cm} copper shielding embedded in another shielding of \unit[15]{cm} of low activity lead.
The inner \unit[5]{cm} of the lead shielding has a specific activity of \unit[$2.7$]{Bq/kg} $^{210}$Pb while the outer \unit[10]{cm} has \unit[$33$]{Bq/kg}.
The spectrometer is located in a measuring chamber which is an additional shielding against radiation from the ambient rock. Furthermore, the detector is constantly held in a nitrogen atmosphere to avoid radon.
The data is collected with a 8192 channel MCA from ORTEC recording energies up to \unit[2.8]{MeV}.  
The full energy peak detection efficiencies are determined with MaGe \cite{Boswell:hc}, a software framework based on Geant4 which was specifically developed for MC simulation of low energy interactions. The simulations are cross checked with an analytically pure SiO$_2$ calibration standard which was measured in the same geometry as the palladium sample. More details can be found in \cite{Lehnert:2011bw,LehnertMaster}. 
%The simulation includes the correct treatment of the angular correlations between \grays\ in the de-excitation \gray\ cascade of the $0^+_1$ and $2^+_2$ final states.

%%%%%%%%%%%%%%%%%%%%%%%%%%%%%%%%%%%%%%%%%%%%%%%%%%%%%%%%%%%%%%%%%%%%%%%%%%%%%%%%%%%%%%%%%%%%%%%%%%%%%%%
\subsection{HADES Dataset}

The measurement was performed in the HADES underground laboratory \cite{HADES:2011} on the premises of the Belgian Nuclear Research Center SCK$\cdot$CEN in Mol, Belgium. The HADES laboratory has an overburden of \unit[500]{m.w.e.}\ reducing the muon flux by about a factor of 5000.
The palladium sample was measured for 44.8 days (\unit[35.9]{kg$\cdot$d} exposure) with a two-detector sandwich setup as illustrated in \fig \ref{pic:PdCombined_detScheme} (b). The bottom detector (Ge6) is a p-type HPGe in a copper cryostat with a relative efficiency of \unit[80]{\%}. The top detector (Ge7) is a n-type HPGe in an aluminium cryostat with a relative efficiency of \unit[90]{\%}. The detectors are surrounded by a shielding consisting of an outer layer of \unit[14.5]{cm} \unit[18]{Bq/kg} (\nuc{Pb}{210}) lead, an intermediate layer of  \unit[4.0]{cm} \unit[2.4]{Bq/kg} (\nuc{Pb}{210}) low activity lead and an inner layer of  \unit[3.5]{cm} electrolytic copper with less than \unit[9]{$\mu$Bq/kg} \nuc{Co}{60} and less than \unit[20]{$\mu$Bq/kg} \nuc{Th}{228} \cite{Wieslander:2009ex}.
The data is collected with a standard GENIE DAQ system in histogram mode for each HPGe detector independently.
An additional custom made list-mode DAQ which also records the detector coincidence and a muon veto signal was not fully operational during the data taking.
The full energy peak detection efficiencies in the setup were determined with the EGS4 software \cite{Lehnert:2013ch,EGS4}.

%%%%%%%%%%%%%%%%%%%%%%%%%%%%%%%%%%%%%%%%%%%%%%%%%%%%%%%%%%%%%%%%%%%%%%%%%%%%%%%%%%%%%%%%%%%%%%%%%%%%%%%
\subsection{LNGS Dataset}

The measurement was performed at the LNGS of INFN in L'Aquila, Italy with an overburden of \unit[3500]{m.w.e.}\ reducing the muon flux by about 6 orders of magnitude.
The palladium sample was measured for \unit[87.2]{d} (\unit[70.0]{$\rm kg\cdot d$} exposure) with a setup of four similar sized {HPGe} detectors as illustrated in \fig \ref{pic:PdCombined_detScheme} (c) with 55 to \unit[57]{\%} relative efficiency. The detectors of \unit[$\approx225$]{cm$^3$} each are installed inside a single cryostat with the endcap facing upwards. The palladium sample is arranged in the central well facing the lateral sides of the detectors. The whole setup is enclosed in a passive shielding made of \unit[25]{cm} low-radioactivity lead as an outer layer and \unit[5]{cm} low-radioactivity copper as an inner layer. In addition the setup is ventilated with nitrogen to remove radon.
The full energy \gray\ detection efficiency of the system is determined with MC simulations based on MaGe \cite{Boswell:hc}. A validation of the simulation has been performed e.g.\ in \cite{Cattadori:2005cfb} and references therein. The data acquisition consists of a four channel ADC system from XIA Inc.\ (Pixie-4), recording the energy of each detector and the time of the event in list mode if a trigger is give by any detector.\\

A feasibility study of a coincidence analysis using multi-detector events to tag the \gray\ cascade 
showed a significantly lower sensitivity for coincidence events compared to the single detector events considering each detector individually. For the \nuc{Pd}{110} $0^+_1$ decay mode, the efficiency to fully detect the \unit[657.8]{keV} or the \unit[815.3]{keV} \gray\ in one detector and triggering one other detector as well is \unit[0.42]{\%}. This can be compared to the efficiency of \unit[2.64]{\%} and \unit[2.30]{\%} to detect these \grays\ in any single detector without triggering another one. The reason is the large self absorption in the sample in combination with the unfavorable geometrical configuration of the source facing the lateral sides of the detectors. This is creating additional attenuation in the crystal holders and is reducing the detection efficiency of \gray\ cascades in multiple detectors. Thus, for this analysis only an anti-coincidence cut was applied between the detectors which reduces the environmental background more than it reduces the detection efficiency.\\

\begin{figure}[h]
\centering
                \includegraphics[width=\textwidth]{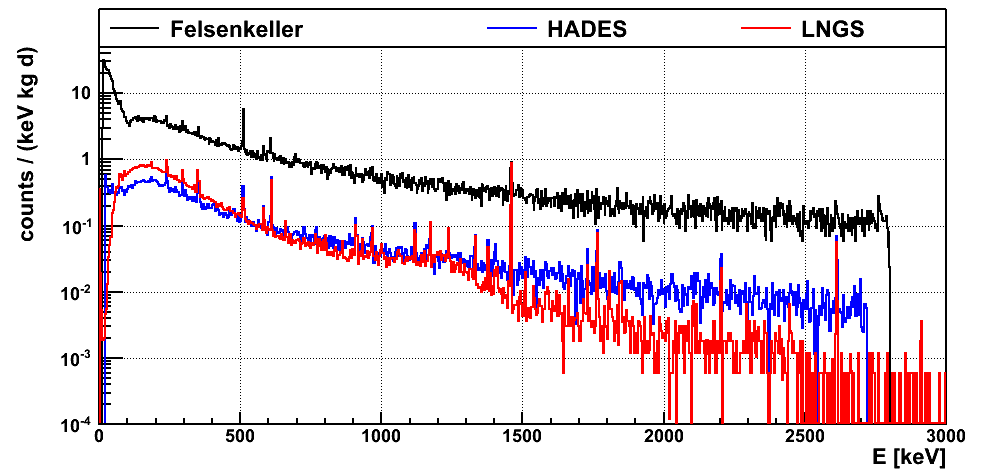}
         \caption{\label{pic:PdESpec} Energy spectra of all three measurements normalized to counts per \unit{keV}, \unit{kg} HPGe detector mass and day. The HADES spectrums shows the combination of detectors Ge6 and Ge7 whereas the LNGS spectrum shows the combination of detectors Ge1, Ge2, Ge3 and Ge4 as described in the text.}
\end{figure}

A comparison of the key parameters of each measurement is shown in \tab \ref{tab:PdMeasurementOverview}. The sample spectrum of each measurement is shown in \fig \ref{pic:PdESpec} normalized to keV, day and kg detector mass. For the HADES and LNGS measurements the combined sum spectra of the detectors are shown. The continuous component in the spectrum, which is produced by muons and neutrons, is about an order of magnitude larger in the Felsenkeller than in HADES due to the smaller overburden. The continous component is further reduced in the LNGS spectrum which lowers the background above the \nuc{K}{40} peak at \unit[1460.8]{keV}. The relative background per detector mass and time in the regions of interest for the de-excitation \grays\ of \nuc{Pd}{110} and \nuc{Pd}{102} is slightly smaller in the LNGS setup compared to the HADES setup. The absolute background in the setup per time is comparable due to the larger amount of detectors. However, due to the unfavorable source-detector geometry, the larger number of detectors does not increase the detection efficiency compared to the HADES setup and thus makes not use of the lower background environment. For this reason the expected sensitivity of the LNGS measurement is comparable with the one from the HADES measurement and all datasets are combined for improved sensitivity.
The radioactivity from the sample itself is subdominant in the overall background for all measurements.

\begin{table}[h]
\begin{center}
\caption{Overview of key parameters of the palladium measurements: the background (bg) and detection efficiency ($\epsilon$) are shown exemplary for the \unit[657.8]{keV} \gline.}
\label{tab:PdMeasurementOverview}
\begin{tabular}{l|ccccl}
\br
location & overburden 				&  time  & bg   & $\epsilon$   & gamma-spectrometry setup          \\
   & [\unit{m}] / [\unit{m.w.e.}] 	&  	[\unit{d}]		& [\unit{cts/keV/d}]  & [\%]	  &     \\
\mr
Felsenkeller	& 	47 / 110				&	16.2				&		1.81	&	3.06			& single HPGe\\
{HADES} 	&	223 / 500			&	44.8				&		0.28			&	4.70  	        & two {HPGe} sandwich\\
{LNGS}	 		& 	1400	 / 3500			& 	87.2				&		0.30	&	2.57			& four {HPGe} setup \\
\br
\end{tabular}\\ 
\end{center}
\end{table}

%%%%%%%%%%%%%%%%%%%%%%%%%%%%%%%%%%%%%%%%%%%%%%%%%%%%%%%%%%%%%%%%%%%%%%%%%%%%%%%%%%%%%%%%%%%%%%%%%%%%%%%
%%%%%%%%%%%%%%%%%%%%%%%%%%%%%%%%%%%%%%%%%%%%%%%%%%%%%%%%%%%%%%%%%%%%%%%%%%%%%%%%%%%%%%%%%%%%%%%%%%%%%%%
%%%%%%%%%%%%%%%%%%%%%%%%%%%%%%%%%%%%%%%%%%%%%%%%%%%%%%%%%%%%%%%%%%%%%%%%%%%%%%%%%%%%%%%%%%%%%%%%%%%%%%%
\section{Analysis}

The analysis is performed on the three datasets $d$ (Felsenkeller, HADES and LNGS). Each de-excitation \gline\ $k$ in a given decay mode has its own fit region $r$ of \unit[$\pm30$]{keV} around the \gline. An exception is the \nuc{Pd}{102} $0^+_1$ mode in which the two de-excitation \glines\ of \unit[468.6]{keV} and \unit[475.1]{keV} are combined into a single fit region. Thus the \nuc{Pd}{110} $2^+_1$, $0^+_1$ and $2^+_2$ transitions have $1$, $2$ and $3$ fit regions in each dataset, respectively. The \nuc{Pd}{102} $2^+_1$, $0^+_1$ and $2^+_2$ transitions have $1$, $1$ and $3$ fit regions, respectively.
The signal count expectation $s_{d,k}$ of each \gline\ in each dataset depends on the half-life $T_{1/2}$ of the decay mode as
\begin{eqnarray}
\label{eq:PdHLtoCounts}
s_{d,k} =
\ln{2} \cdot  \frac{1}{T_{1/2}} \cdot \epsilon_{d,k} \cdot N_A \cdot t_d \cdot m \cdot f_{\rm iso}  \cdot \frac{1}{M_{\rm Pd}}\ ,
\end{eqnarray}

where $\epsilon_{d,k}$ is the full energy detection efficiency of \gline\ $k$ in dataset $d$, $
N_A$ is the Avogadro constant,
$t_d$ is the live-time of the dataset, 
$m$ is the mass of the palladium sample,
$M_{\mathrm{Pd}}$ the molar mass and
$f_{\rm iso}$ is the isotopic natural abundance of \nuc{Pd}{102} and \nuc{Pd}{110}, respectively. The data is binned with \unit[0.68]{keV}, \unit[0.5]{keV}, \unit[1.0]{keV} wide bins for the Felsenkeller, HADES and LNGS datasets respectively.
The Bayesian Analysis Toolkit (BAT) \cite{Caldwell:2009kh} is used to perform a maximum posterior fit combining all three datasets and \glines\ for a given decay mode. The likelihood $\mathcal{L}$ is defined as the product of the Poisson probabilities of each bin $i$ in fit region $r$ in every dataset $d$ 
\begin{eqnarray}
\mathcal{L}(\mathbf{p}|\mathbf{n}) =
\prod \limits_d \prod \limits_r \prod \limits_i \frac{\lambda_{d,r,i}(\mathbf{p})^{n_{d,r,i}}}{n_{d,r,i}!} e^{-\lambda_{d,r,i}(\mathbf{p})}\ ,
\end{eqnarray}

where \textbf{n} denotes the data and \textbf{p} the set of floating parameters.
$n_{d,r,i}$ is the measured number of counts and $\lambda_{d,r,i}$ is the expected number of counts in bin $i$. $\lambda_{d,r,i}$ is taken as the integral of the extended p.d.f.\ $P_{d,r}$ in this bin.
%
%\begin{eqnarray}
%\lambda_{d,r,i}(\mathbf{p}) &=&
% \int_{\Delta E_{d,r,i}} P_{d,r}(E|\mathbf{p}) dE\ , \label{eq:Pd_expcounts}
%\end{eqnarray}
%
%where $\Delta E_{d,r,i}$ is the bin width. 
The counts in the fit region are expected from three different types of contributions which are used to construct $P_{d,r}$:
(1) A linear background, (2) the Gaussian signal peak and (3) a number of Gaussian background peaks. The number and type of background peaks depend on the fit region and will be described later. The full expression of $P_{d,r}$ is written as:
\begin{eqnarray}
P_{d,r}(E|\mathbf{p}) &&=
 B_{d,r} + C_{d,r}\left( E-E_0 \right) \label{eq:Pd_pdf}\\[2mm]
&&+  \frac{s_{d,k}}{\sqrt{2\pi}\sigma_{d,k}} 
\cdot \exp{\left(-\frac{(E-E_{k})^2}{2\sigma_{d,k}^2}\right)}\nonumber\\[2mm]
&&+ \sum \limits_{l_r} \left[\frac{b_{d,l_r}}{\sqrt{2\pi}\sigma_{d,k}} 
\cdot \exp{\left(-\frac{(E-E_{l_r})^2}{2\sigma_{d,k}^2}\right)}\right].\nonumber
%+ \left(B_{0,d} + B_{1,d}E_{d,i}\right). \label{eqn:pdf}
\end{eqnarray}

The first line is describing the linear background with the two parameters $B_{d,r}$\ and $C_{d,r}$.
The second line is describing the signal peak with the energy resolution $\sigma_{d,k}$ and the \gline\ energy $E_k$.
The third line is describing the $l_r$ background peaks in fit region $r$ with the strength of the peak $b_{d,l_r}$ and the peak position $E_{l_r} $. The same p.d.f.\ with different parameter values is used for all three datasets. Hence, the same number of background peaks is used in every dataset even if not all background peaks are prominent in every datasets.\\

The free parameters $\mathbf{p}$ in the fit are: 
\begin{itemize}
\item 1 inverse half-life $(T_{1/2})^{-1}$ with flat prior
\item 2 x $3$ x $r$ linear background parameters $B_{d,r}$ and $C_{d,r}$ with flat priors
\item $3$ x $k$ energy resolutions $\sigma_{d,k}$ with Gaussian priors
\item $3$ x $k$ detection efficiencies $\epsilon_{d,k}$ with Gaussian priors
\item $k$ signal peak positions $E_{k}$ with Gaussian priors
\item $3$ x $l_r$ x $r$ background peak strength $b_{d,l_r}$ with flat priors 
\item $l_r$ x $r$ background peak positions $E_{l_r}$ with Gaussian priors
\end{itemize}

In total this amounts to 30 fit parameters for the $2^+_1$ mode, 59 parameters for the $0^+_1$ mode and 76 parameters for the $2^+_2$ mode of \nuc{Pd}{110} and 18 parameters for the $2^+_1$ mode, 25 parameters for the $0^+_1$ mode and 52 parameters for the $2^+_2$ mode of \nuc{Pd}{102}.\\

The energy resolutions are determined with calibration spectra for each dataset independently. The mean of the Gaussian priors is taken from these calibrations and reported in \tab \ref{tab:PdMeasurementSigma} in the appendix. The width of these priors is taken as the uncertainty of the resolution calibration curve and approximated with \unit[5]{\%} for all datasets and \glines.\\

The detection efficiencies are determined with MC simulations for each dataset and transition independently. 
The mean value of the Gaussian prior is reported in \tab \ref{tab:PdMeasurementEfficiencies} in the appendix. 
For the HADES dataset the efficiencies are taken from \cite{Lehnert:2013ch}. For the Felsenkeller dataset the efficiencies are reevaluated compared to \cite{Lehnert:2011bw} to include the $2^+_2$ transitions. 
The uncertainty of the detection efficiencies is approximated with \unit[10]{\%} for each dataset and \gline\ and used for the width of the prior. The mean values and uncertainties of the signal and background peak positions are taken from nuclear data sheets \cite{website:NuDat2}.\\

The posterior probability is calculated from the likelihood and prior probabilities with BAT. The maximum of the posterior probability is the best fit. The marginalized posterior probability distribution of $(T_{1/2})^{-1}$ is extracted and the \unit[90]{\%} quantile of this distribution is used for setting a \unit[90]{\%} credibility limit on the half-life. Systematic uncertainties are included via the distribution of the priors. The influence of the systematic uncertainties on the half-life is \unit[$<2$]{\%} which is evaluated by fixing all Gaussian priors to their mean value and repeating the analysis.

%%%%%%%%%%%%%%%%%%%%%%%%%%%%%%%%%%%%%%%%%%%%%%%%%%%%%%%%%%%%%%%%%%%%%%%%%%%%%%%%%%%%%%%%%%%%%%%%%%%%%%%
%%%%%%%%%%%%%%%%%%%%%%%%%%%%%%%%%%%%%%%%%%%%%%%%%%%%%%%%%%%%%%%%%%%%%%%%%%%%%%%%%%%%%%%%%%%%%%%%%%%%%%%
%%%%%%%%%%%%%%%%%%%%%%%%%%%%%%%%%%%%%%%%%%%%%%%%%%%%%%%%%%%%%%%%%%%%%%%%%%%%%%%%%%%%%%%%%%%%%%%%%%%%%%%

\section{Results}
\label{sec:results}

The analysis of each decay mode is discussed for \nuc{Pd}{110} and \nuc{Pd}{102}. No signal is observed for either decay mode or nuclide and half-life limits are extracted.

%#####################################################################################################
%#####################################################################################################
\subsection{\nuc{Pd}{110} Decay Mode $0\nu[2\nu]\beta\beta\ 0^+_{\rm g.s.} - 2^+_1$}

\begin{figure}[h]
\centering
\includegraphics[width=\textwidth]{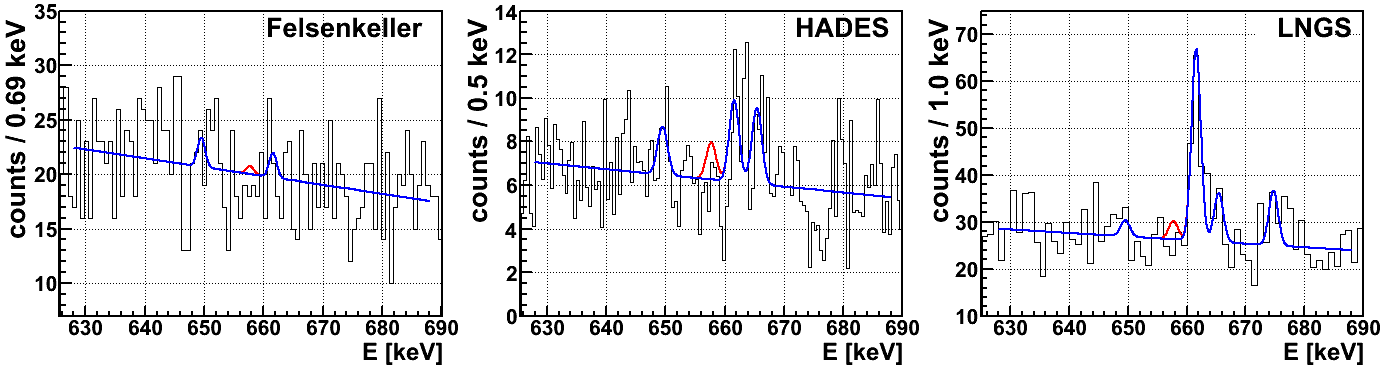}
         \caption{\label{pic:pdf_ROI_Composition_Pd110_2p1} Fit regions for the \nuc{Pd}{110} $2^+_1$ decay mode for all datasets. Shown is the best fit in blue and the best fit with the signal strength set to the \unit[90]{\%} C.I.\ half-life limit in red. Note that each dataset has a different binning and measuring time.}
\end{figure}

The fit region of \unit[$\pm30$]{keV} is centered around the single \gline\ of \unit[657.8]{keV} and illustrated in \fig \ref{pic:pdf_ROI_Composition_Pd110_2p1} for all three datasets. 
Three known background \glines\ from decay chain nuclides are included in the fit coming from \nuc{Bi}{210m} at \unit[649.6]{keV} (\unit[3.4]{\%}), \nuc{Bi}{214} at \unit[665.5]{keV} (\unit[1.5]{\%}) and from \nuc{Ac}{228} at \unit[674.8]{keV} (\unit[2.1]{\%}), where the value in parentheses is the emission probability of this \gray. Additionally, \nuc{Cs}{137} at \unit[661.7]{keV} (\unit[85.1]{\%}) is included in the fit as an anthropogenic background. This contamination can be clearly seen in the HADES and LNGS datasets. In comparison, the background peaks in the Felsenkeller dataset are not significant. Here, the background is dominated by atmospheric muons due to the smaller overburden.\\

The best fit is shown as the blue p.d.f.\ in \fig \ref{pic:pdf_ROI_Composition_Pd110_2p1}. The signal peak according to the \unit[90]{\%} C.I.\ is shown as the red p.d.f.\ The best fit yields zero expected events from the signal process. The obtained \unit[90]{\%} quantile of the posterior translates into a half-life limit for the \nuc{Pd}{110} $2^+_1$ decay mode of
\begin{equation}
T_{1/2} > 2.9 \cdot 10^{20}\, {\rm yr}\ (\unit[90]{\%}\ \rm C.I.)\ .
\end{equation}

%The size of the signal peak at \unit[90]{\%} C.I.\ illustrates the sensitivity of the dataset. The Felsenkeller measurement with a larger background, less measuring time and a single detector setup with smaller full energy detection efficiency has a smaller sensitivity. The HADES and LNGS datasets show roughly the same sensitivity.

%#####################################################################################################
%#####################################################################################################
\subsection{\nuc{Pd}{110} Decay Mode $0\nu[2\nu]\beta\beta\ 0^+_{\rm g.s.} - 0^+_1$}

\begin{figure}[h]
\centering
                \includegraphics[width=\textwidth]{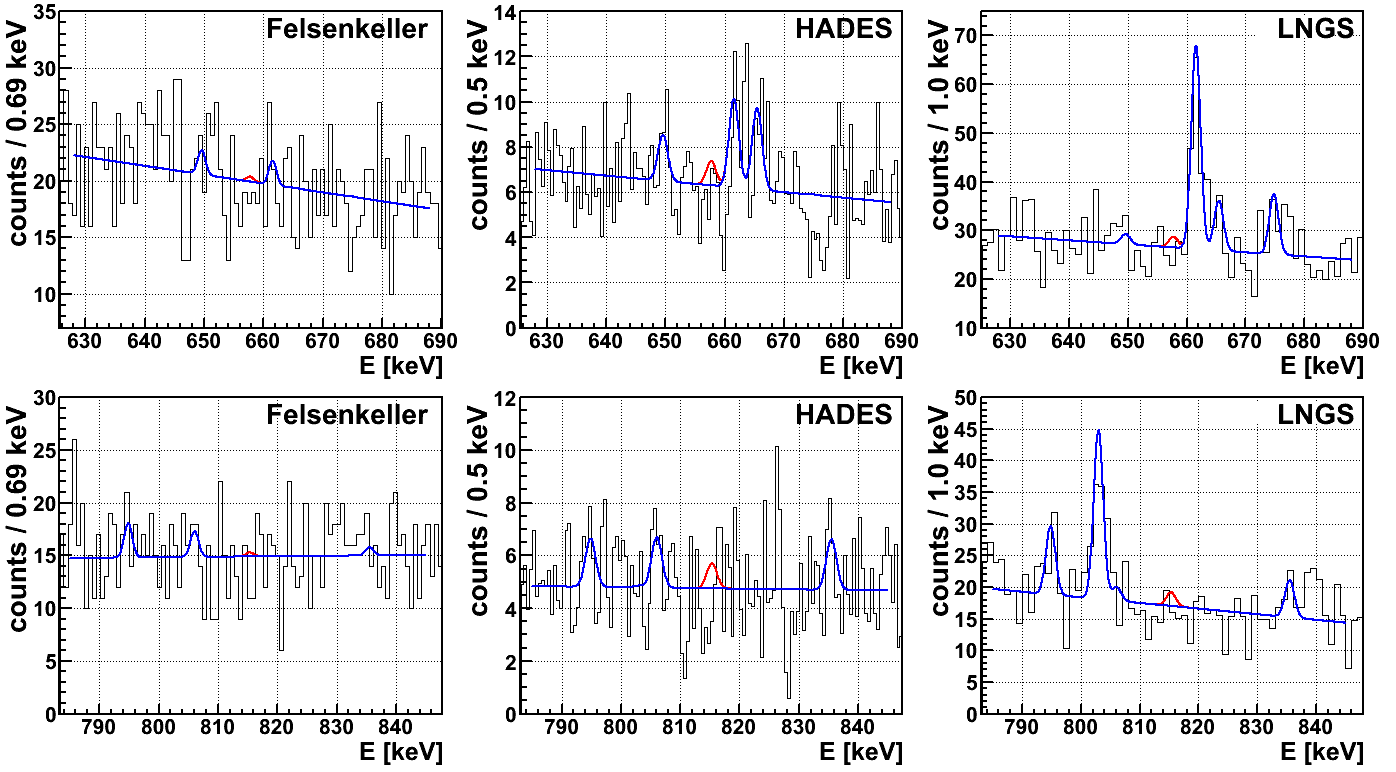}
         \caption{\label{pic:pdf_ROI_Composition_Pd110_0p1} Fit regions for the \nuc{Pd}{110} $0^+_1$ decay mode for all datasets. Shown is the best fit in blue and the best fit with the signal strength set to the \unit[90]{\%} C.I.\ half-life limit in red. Note that each dataset has a different binning and measuring time.}
\end{figure}

For this decay mode two fit regions are selected. The first is the same as for the $2^+_1$ transition and the second is centered around the de-excitation \gline\ of \unit[815.3]{keV}. The fit regions are shown in \fig \ref{pic:pdf_ROI_Composition_Pd110_0p1}.\\

%Both \grays\ occur in coincidence with an angular correlation ($W(\theta) \propto 1-3\cos^2{\theta} + 4\cos^4{\theta}$ \cite{Schatz:1997wp}) which is considered in the MC simulations using DECAY0. 

Three known background \glines\ are included in the second fit region coming from \nuc{Ac}{228} at \unit[795.0]{keV} (\unit[4.3]{\%}) and at \unit[835.7]{keV} (\unit[1.6]{\%}) and from \nuc{Bi}{214} at \unit[806.1]{keV} (\unit[1.3]{\%}). 
%
%
%Another prominent \gline\ is visible in the LNGS dataset
%at \unit[803.1]{keV} which is coming from the decay of \nuc{210}{Po} in
%the lead shield. It is included in the fit as a background
%\gline.
Another prominent \gline\ is visible in the LNGS dataset at \unit[803.1]{keV} which is potentially coming from the first excited state in \nuc{Pb}{206} after excitation with elastic neutron scattering ($n$,$n'$). It is included in the fit as a background \gline.
Yet another potential \gline\ can be seen at \unit[826.5]{keV} in the HADES dataset. This \gline\ could not be identified and is therefore not included as a background. However, including this \gline\ in a test fit changes the combined half-life limit by only \unit[2.2]{\%} compared to not including it in the fit.\\

The final half-life limit for the \nuc{Pd}{110} $0^+_1$ decay mode is
\begin{equation}
T_{1/2} > 4.0 \cdot 10^{20}\, {\rm yr}\ (\unit[90]{\%}\ \rm C.I.)\ .
\end{equation}

%#####################################################################################################
%#####################################################################################################
\subsection{\nuc{Pd}{110} Decay Mode $0\nu[2\nu]\beta\beta\ 0^+_{\rm g.s.} - 2^+_2$}

\begin{figure}[h]
\centering
                \includegraphics[width=\textwidth]{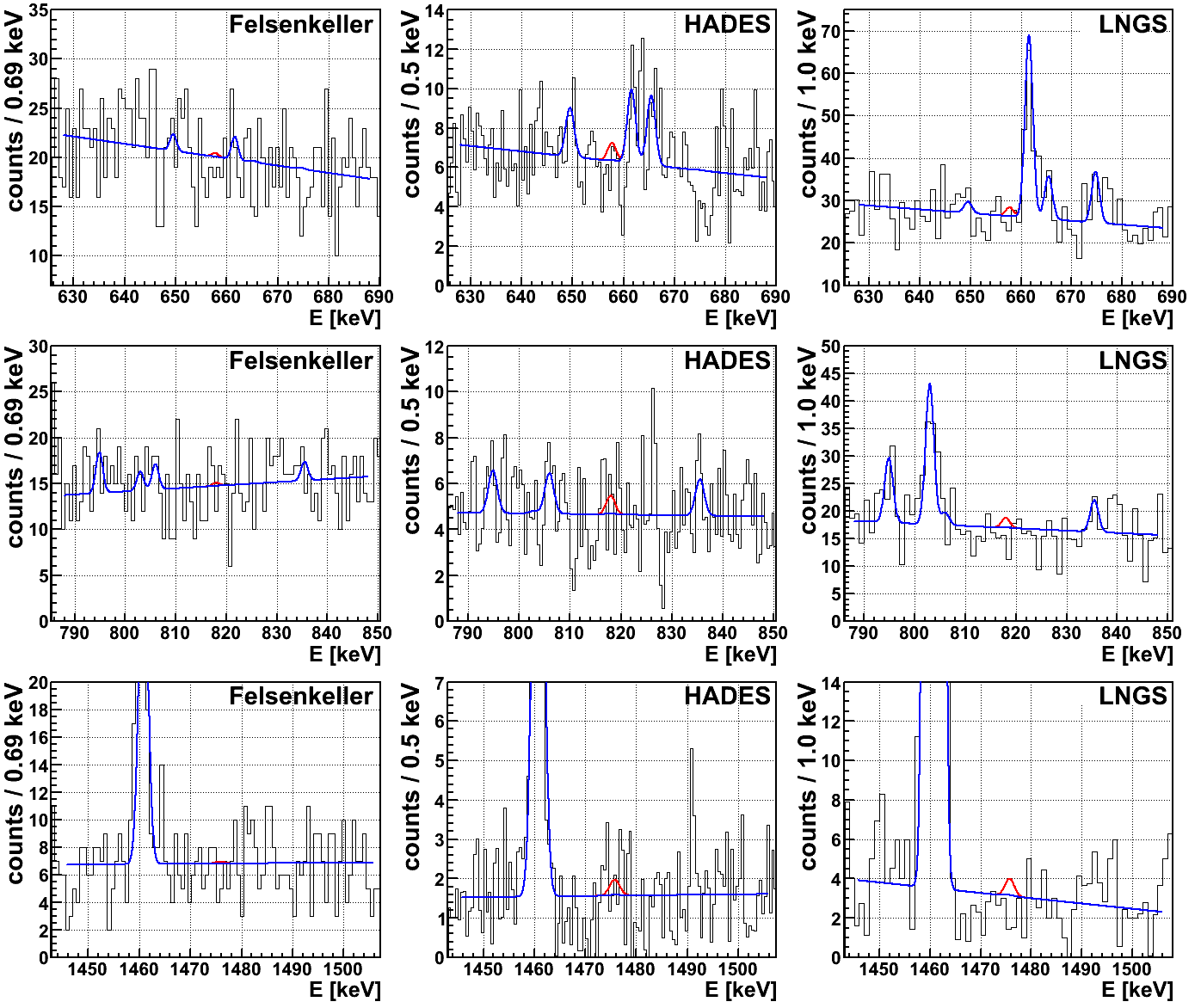}
         \caption{\label{pic:pdf_ROI_Composition_Pd110_2p2} Fit regions for the \nuc{Pd}{110} $2^+_2$ decay mode for all datasets. Shown is the best fit in blue and the best fit with the signal strength set to the \unit[90]{\%} C.I.\ half-life limit in red. Note that each dataset has a different binning and measuring time.}
\end{figure}

This decay mode has two decay branches, one with a single \gray\ emission (\unit[35.5]{\%}) and one with two coincident \grays\ (\unit[64.5]{\%}). 
%
%The two \grays\ are emitted with an angular correlation ($W(\theta) \propto  1-\frac{15}{13}\cos^2{\theta} + \frac{16}{13}\cos^4{\theta}$ \cite{Schatz:1997wp}) which is included in the MC via DECAY0. 
%
In total three fit regions are selected. The first is the same as for the $2^+_1$ transition. The second region is centered around the \unit[818.0]{keV} \gline\ and the third region is centered around the \unit[1475.8]{keV} \gline. All fit regions are shown in \fig \ref{pic:pdf_ROI_Composition_Pd110_2p2}.\\ 

In the first and second fit region the background \glines\ are included as described in the other transitions above. In the third fit region the \nuc{K}{40} \gline\ at \unit[1460.83]{keV} (\unit[10.7]{\%}) is the most prominent feature in all datasets and is included in the fit. Another potential \gline\ at \unit[1490.7]{keV} can be seen in the HADES dataset with \unit[1.8]{$\sigma$} significance. Some indication can also be seen in the LNGS dataset with \unit[1.3]{$\sigma$}. This potential \gline, however, could not be identified and is therefore not included in the fit. The difference in the half-life limit between including and not including the potential \glines\ at \unit[826.5]{keV} and \unit[1490.7]{keV} is \unit[3]{\%} and thus not very strong.\\

The final half-life for the \nuc{Pd}{110} $2^+_2$ decay mode is
\begin{equation}
T_{1/2} > 3.0 \cdot 10^{20}\, {\rm yr}\ (\unit[90]{\%}\ \rm C.I.)\ .
\end{equation}

%#####################################################################################################
%#####################################################################################################
\subsection{\nuc{Pd}{102} Decay Mode $0\nu[2\nu]\beta\beta\ 0^+_{\rm g.s.} - 2^+_1$}
\begin{figure}[h]
\centering
                \includegraphics[width=\textwidth]{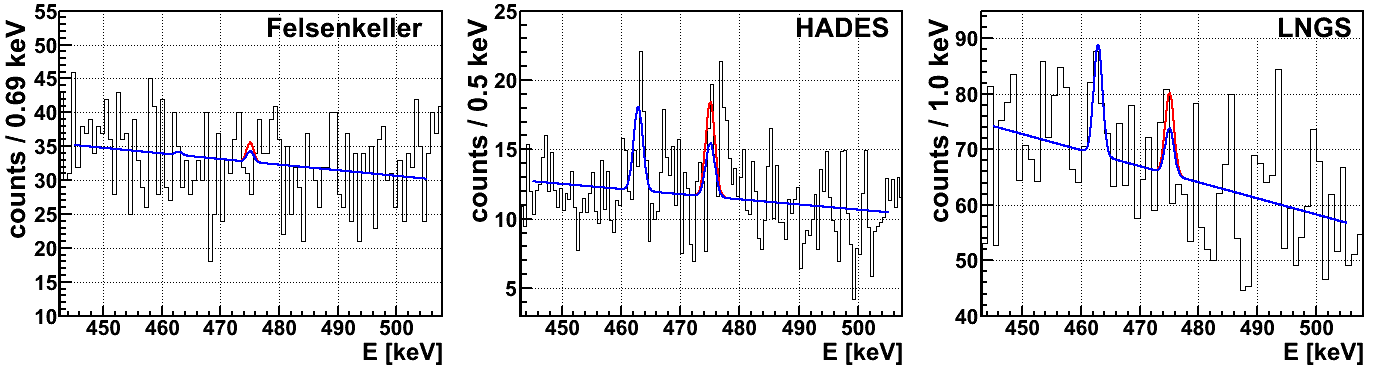}
         \caption{\label{pic:pdf_ROI_Composition_Pd102_2p1} Fit regions for the \nuc{Pd}{102} $2^+_1$ decay mode for all datasets. Shown is the best fit in blue and the best fit with the signal strength set to the \unit[90]{\%} C.I.\ half-life limit in red. Note that each dataset has a different binning and measuring time.}
\end{figure}

The single fit region of the \nuc{Pd}{102} $2^+_1$ transition is centered \unit[$\pm$]{30} keV around the \unit[475.1]{keV} de-excitation \gray\ energy and is shown in \fig \ref{pic:pdf_ROI_Composition_Pd102_2p1}.\\

The \nuc{Ac}{228} background \gline\ at \unit[463.0]{keV} (\unit[4.4]{\%}) is included in the fit. Some potential peak structures can be seen in the HADES spectrum close to or directly underneath the signal peak. They cannot be clearly identified and appear to be at \unit[$\approx475$]{keV} and \unit[$\approx477$]{keV}. A search for radionuclides with \gray\ emission at those energies as well as a search for potential \gray\ summation or escape effects remained inconclusive. 
%Another investigation was performed looking at the two detectors in the HADES measurements individually. 
The peak structures are only visible in the bottom detector Ge6 with copper endcap and \unit[0.9]{mm} dead layer. The top detector Ge7 with Al endcap and \unit[0.3]{micron} dead layer, does not show these features. They are also not visible in the background spectrum of the setup. Hence, those \gline\ features are either a background fluctuation in Ge6 alone or an unknown irreducible background contribution. In both cases they cannot be included in the fit.\\

Performing the fit on all three datasets results in a best fit and limit as shown in \fig \ref{pic:pdf_ROI_Composition_Pd102_2p1}. The best fit finds a half-life of \unit[\baseT{1.2}{19}]{yr} which includes the no signal case in the smallest connected \unit[98.4]{\%} or \unit[2.41]{$\sigma$} interval. The significance of the signal peak is almost entirely due to the background features in the HADES dataset. Performing the same fit only on the Felsenkeller and LNGS datasets results in a best fit consistent with zero at the \unit[0.77]{$\sigma$} level. The \unit[90]{\%} C.I.\ lower limit with the HADES dataset is \unit[\baseT{7.6}{18}]{yr} compared to \unit[\baseT{7.2}{18}]{yr} without the HADES dataset. With the HADES dataset the limit is reduced due to the upwards fluctuations of background in the peak region whereas without the HADES dataset the sensitivity is smaller.\\

For the final limit of the \nuc{Pd}{102} $2^+_1$ transition the HADES dataset is included due to the larger sensitivity. The limit is set to 
\begin{equation}
T_{1/2} > 7.6 \cdot 10^{18}\, {\rm yr}\ (\unit[90]{\%}\ \rm C.I.)\ .
\end{equation}

%#####################################################################################################
%#####################################################################################################
\subsection{\nuc{Pd}{102} Decay Mode $0\nu[2\nu]\beta\beta\ 0^+_{\rm g.s.} - 0^+_1$}

\begin{figure}[h]
\centering
                \includegraphics[width=\textwidth]{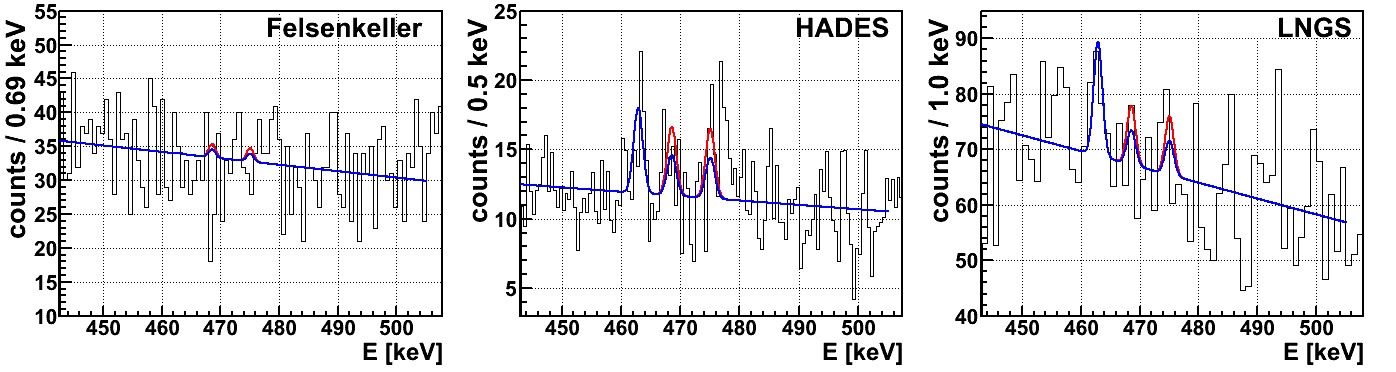}
         \caption{\label{pic:pdf_ROI_Composition_Pd102_0p1} Fit regions for the \nuc{Pd}{102} $0^+_1$ decay mode for all datasets. Shown is the best fit in blue and the best fit with the signal strength set to the \unit[90]{\%} C.I.\ half-life limit in red. Note that two signal peaks are included in the same fit region.}
\end{figure}

This decay mode has two coincident \gray\ emissions at \unit[475.1]{keV} and \unit[468.6]{keV} which are analyzed in a single fit region identical to the one for the $2^+_1$ transition above. This is the only decay mode with two signal peaks in the same fit region which adds another signal term to the p.d.f.\ in \eq \ref{eq:Pd_pdf}. The fit is shown in \fig \ref{pic:pdf_ROI_Composition_Pd102_0p1} using the same background \gline\ from \nuc{Ac}{228} at \unit[463.0]{keV} as before.\\

Also in this case the unidentified peak features are present and are treated in the same way as above. In addition there is a small upwards fluctuation underneath the \unit[468.6]{keV} signal peak in the HADES dataset as well. Also this fluctuation is only visible in one of the HADES detectors and is thus treated as a background fluctuation. With the HADES dataset the smallest interval containing the no signal case is \unit[98.7]{\%} or \unit[2.49]{$\sigma$}. Excluding the HADES dataset results in a \unit[0.1]{$\sigma$} difference to the no signal case. The difference in half-life limit is \unit[\baseT{8.8}{18}]{yr} including HADES and \unit[\baseT{1.0}{19}]{yr} excluding HADES.\\

The final limit for the \nuc{Pd}{102} $0^+_1$ transition is taken as the one with the HADES dataset and set to 
\begin{equation}
T_{1/2} > 8.8 \cdot 10^{18}\, {\rm yr}\ (\unit[90]{\%}\ \rm C.I.)\ .
\end{equation}

%#####################################################################################################
%#####################################################################################################
\subsection{\nuc{Pd}{102} Decay Mode $0\nu[2\nu]\beta\beta\ 0^+_{\rm g.s.} - 2^+_2$}

\begin{figure}[h]
\centering
                \includegraphics[width=\textwidth]{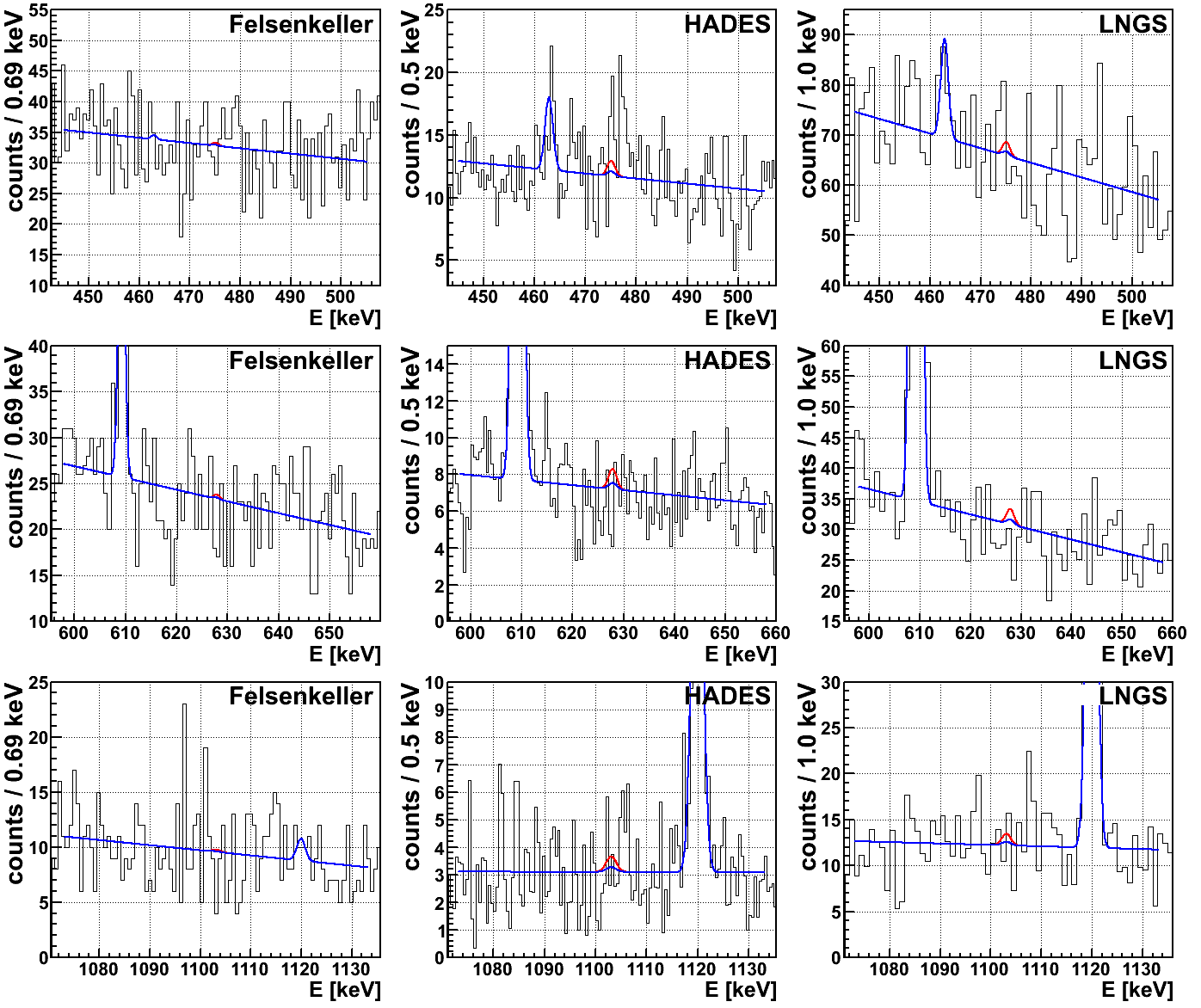}
         \caption{\label{pic:pdf_ROI_Composition_Pd102_2p2} Fit regions for the \nuc{Pd}{102} $2^+_2$ decay mode for all datasets. Shown is the best fit in blue and the best fit with the signal strength set to the \unit[90]{\%} C.I.\ half-life limit in red. Note that each dataset has a different binning and measuring time.}
\end{figure}

This decay mode has two decay branches with a coincident double \gray\ emission of \unit[475.1]{keV} and \unit[627.9]{keV} (\unit[62.9]{\%}) and a single \gray\ emission of \unit[1103.1]{keV} (\unit[37.1]{\%}). The first fit region is identical to the one in the $2^+_1$ transition. The second fit region is centered around the \unit[627.9]{keV} \gline\ and includes the \nuc{Bi}{214} background \gline\ at \unit[609.3]{keV} (\unit[45.5]{\%}). The third fit region is centered around the \unit[1103.1]{keV} \gline\ and includes the \nuc{Bi}{214} background \gline\ at \unit[1120.3]{keV} (\unit[14.9]{\%}). All three regions including the fit are shown in \fig \ref{pic:pdf_ROI_Composition_Pd102_2p2}.\\

In this case, the background fluctuations in the first fit region of the HADES dataset do not have a strong influence since the fit is dominated by the other two fit regions in which no fluctuations occur. The final limit for the \nuc{Pd}{102} $2^+_2$ transition is found as
\begin{equation}
T_{1/2} > 1.4 \cdot 10^{19}\, {\rm yr}\ (\unit[90]{\%}\ \rm C.I.)\ .
\end{equation}

%%%%%%%%%%%%%%%%%%%%%%%%%%%%%%%%%%%%%%%%%%%%%%%%%%%%%%%%%%%%%%%%%%%%%%%%%%%%%%%%%%%%%%%%%%%%%%%%%%%%%%%
%%%%%%%%%%%%%%%%%%%%%%%%%%%%%%%%%%%%%%%%%%%%%%%%%%%%%%%%%%%%%%%%%%%%%%%%%%%%%%%%%%%%%%%%%%%%%%%%%%%%%%%
%%%%%%%%%%%%%%%%%%%%%%%%%%%%%%%%%%%%%%%%%%%%%%%%%%%%%%%%%%%%%%%%%%%%%%%%%%%%%%%%%%%%%%%%%%%%%%%%%%%%%%%

\section{Discussion and Conclusions}

Excited state transitions in the double beta decay candidate nuclides of \nuc{Pd}{110} and \nuc{Pd}{102} were investigated. A new measurement was performed at LNGS in a deeper location with lower ambient background per detector mass and with a larger 4-detector setup compared to previous measurements. However, this measurement by itself could not significantly improve the previous limit due to an unfavorable detectors-sample configuration reducing the detection efficiency. Thus the data was combined with the two previous measurements at the Felsenkeller and HADES underground laboratories. This combined analysis features the following improvements:

\begin{itemize}
\item All available data of the palladium sample was combined improving the overall sensitivity compared to individual datasets.
\item The analysis combines the information of all de-excitation \grays\ in a given decay mode compared to using only a single \gray\ with the best limit in the past analyses. This is improving the sensitivity for the $0^+_1$ and $2^+_2$ excited state transitions with multiple \gray\ emission and multiple decay branches. 
\item The analysis is based on spectral fits compared to simple counting limits in the past. The change of methodology is considered an improvement using the full spectral information; however the influence on the limit setting is expected to be only marginal.
\item The Bayesian half-life limits are set with the full extraction of the posterior probability which naturally includes systematic uncertainties as prior information in the fit and thus makes the results more robust. 
\end{itemize}

The analysis finds no signal for any decay mode and \unit[90]{\%} credibility lower half-life bounds are obtained which are summarized in \tab \ref{tab:PdResults}. The bounds include systematic uncertainties on the efficiency, energy resolution and peak position. The limits for the \nuc{Pd}{102} decay modes are roughly one order of magnitude weaker than for the \nuc{Pd}{110} decay modes due to the smaller isotopic abundance. The previous best limits could be improved by factors of 1.3 to 3 depending on the decay mode. The current experimental sensitivity is still three orders of magnitude smaller than the lowest half-life prediction for the \nuc{Pd}{110} $0^+_1$ transition by the QRPA model \cite{Suhonen:2011du}.\\

\begin{table}[t]
\begin{center}
\caption{Summary of measured half-life limits for investigated \nuc{Pd}{110} and \nuc{Pd}{102} double beta decay excited state transitions.}
\begin{tabular}{cc||cc}
\br
decay mode & $T_{1/2}$ [yr]  &  decay mode &$T_{1/2}$ [yr]\\
		   &  \unit[90]{\%} C.I. &  &\unit[90]{\%} C.I.\\

\mr
\nuc{Pd}{110} $2^+_1$ & \baseT{2.9}{20} &   \nuc{Pd}{102} $2^+_1$ & \baseT{7.6}{18}  \\
\nuc{Pd}{110} $0^+_1$ & \baseT{4.0}{20} &   \nuc{Pd}{102} $0^+_1$ & \baseT{8.8}{18}  \\
\nuc{Pd}{110} $2^+_2$ & \baseT{3.0}{20} &   \nuc{Pd}{102} $2^+_2$ & \baseT{1.4}{19}  \\
\br
\end{tabular}\\ 
\label{tab:PdResults}
\end{center}
\end{table}

% The improvements are mainly due to the combination of three datasets including the new LNGS dataset. The spectral fit and the combination of different fit regions in a single fit especially improves the limits for the higher excited states with multiple \gray\ emission and multiple branching ratios.\\ 
%

Potential future improvements could be achieved by lowering the radioactive background of ambient sources ($B$), increasing the detection efficiency ($\epsilon$) by rearranging the detector setup and in general by using more target material ($m$) and increasing the measuring time ($T$). However, considering the figure of merit for the half-life sensitivity of such an experiment $T_{1/2} \propto \epsilon \cdot \sqrt{m\cdot T / B}$ it is clear that the gap between current sensitivity and the lowest theoretical predictions cannot be bridged by standard gamma-spectroscopy setups. A dedicated experiment would need to reduce or discriminate the ambient background by at least a factor of 10. Since the intrinsic background in the palladium is subdominant and other dedicated double beta decay experiments have proven very low background environments this is not unfeasible. An optimized HPGe detector setup and sample volume would need to increase the detection efficiency by a factor of 10. This could be achieved with a thin layer of palladium between multiple sandwich detectors reducing self-absorption and increasing the solid angle coverage. With an increase in measuring time by a factor of 10 to about \unit[2]{yr}, this would lead to a two orders of magnitude larger half-life sensitivity. Such an arrangement would also allow for a coincidence analysis further reducing the background while remain with a large detection efficiency. To improve by another order of magnitude, the target could be increased in mass and possibly enriched which would, however, dominate the cost for searches in palladium. An alternative option would be to re-use the existing low background environment of current generation double beta decay experiments with good gamma-ray discrimination capability such as e.g.\ GERDA, CUORE or C0BRA.

%\section{Acknowledgements}
%
%\missing Felsenkeller, HADES, LNGS staff?

\bigskip

\section*{References}

%\begin{thebibliography}{10}

\bibliographystyle{unsrt}    
\bibliography{./PdCombined_bib1} 

%\end{thebibliography}

\section{Appendix}

\begin{table}[h]
\centering
\caption{\label{tab:PdMeasurementSigma} Energy resolution in $\sigma_E$ for all peaks and datasets. For the HADES and LNGS measurements the quoted values are the average of all detectors. The uncertainty is approximated with \unit[5]{\%}.}
\begin{tabular}{l|rrr}
\br
\gline\ energy  & Felsenkeller & HADES & LNGS   \\
\mr
%\multicolumn{ 3}{l}{\nuc{Pd}{110}}\\
%\hline

					   \unit[657.8]{keV}  & \unit[0.642]{keV} &  \unit[0.828]{keV} & \unit[0.768]{keV} \\
					   \unit[815.3]{keV}  & \unit[0.683]{keV} &  \unit[0.868]{keV} & \unit[0.794]{keV} \\
					   \unit[818.0]{keV}  & \unit[0.684]{keV} &  \unit[0.868]{keV} & \unit[0.794]{keV} \\
					   \unit[1475.8]{keV} & \unit[0.837]{keV} &  \unit[1.003]{keV}  & \unit[0.912]{keV} \\
\mr
					   \unit[468.6]{keV}  & \unit[0.590]{keV} &  \unit[0.773]{keV} & \unit[0.739]{keV} \\
					   \unit[475.1]{keV}  & \unit[0.592]{keV} &  \unit[0.775]{keV} & \unit[0.740]{keV} \\
					   \unit[627.9]{keV}  & \unit[0.634]{keV} &  \unit[0.819]{keV}  & \unit[0.764]{keV} \\
					   \unit[1103.1]{keV} & \unit[0.754]{keV} &  \unit[0.932]{keV}  & \unit[0.844]{keV} \\
\br
\end{tabular}\\ 
\end{table}

\begin{table}[h]
\centering\caption{\label{tab:PdMeasurementEfficiencies} Detection efficiencies for all \glines\ in all decay modes for all datasets. Values for the HADES setup are taken from \cite{Lehnert:2013ch}. The uncertainty is approximated with \unit[10]{\%}. Note that branching ratios and summation effects change the full energy detection efficiency of a \gline\ in different decay modes.}
\begin{tabular}{rrrr}
\br
 \gline\ energy [keV]  & Felsenkeller & HADES & LNGS\\
\mr
%\multicolumn{ 3}{l}{\nuc{Pd}{110}}\\
%\hline
\multicolumn{ 4}{l}{\nuc{Pd}{110} $2^+_1$ \unit[657.8]{keV} decay mode:}\\
				\unit[657.7]{keV} (\unit[100.0]{\%}) &  \unit[3.06]{\%} & \unit[4.70]{\%} & \unit[2.57]{\%} \\
\mr
\multicolumn{ 4}{l}{\nuc{Pd}{110} $0^+_1$ \unit[1473.1]{keV} decay mode:}\\
				\unit[657.8]{keV} (\unit[100.0]{\%}) & \unit[2.56]{\%}  & \unit[3.94]{\%} & \unit[2.08]{\%} \\
				\unit[815.3]{keV} (\unit[100.0]{\%}) & \unit[2.30]{\%}  & \unit[3.68]{\%}$^a$ & \unit[1.97]{\%} \\
\mr
\multicolumn{ 4}{l}{\nuc{Pd}{110} $2^+_2$ \unit[1475.80]{keV} decay mode:}\\
			    \unit[657.8]{keV}  (\unit[64.5]{\%}) & \unit[1.68]{\%} &  \unit[2.53]{\%} & \unit[1.38]{\%} \\
				\unit[818.0]{keV}  (\unit[64.5]{\%}) & \unit[1.53]{\%} &  \unit[2.40]{\%} & \unit[1.28]{\%} \\
 				\unit[1475.8]{keV} (\unit[35.5]{\%}) & \unit[0.87]{\%} &  \unit[1.32]{\%} & \unit[0.69]{\%}   \\
%\midrule
%\nuc{Pd}{110} $0^+_2$ \unit[1731.33]{keV} \
%				\unit[1073.7]{keV} & \unit[86.73]{\%} &  \unit[1.89]{\%} & 10.1 & \baseT{8.50}{19} \\
%				\unit[657.76]{keV}\footnotemark[1] & \unit[95.32]{\%} &  \unit[3.78]{\%} & 12.4 & \baseT{1.38}{20} \\
%				\unit[255.49]{keV} & \unit[13.27]{\%} &  \unit[0.36]{\%} & 25.3 &\baseT{6.46}{18} \\
%				\unit[1475.80]{keV} & \unit[4.68]{\%} &  \unit[0.12]{\%} & 11.5 &\baseT{4.87}{18} \\
%				\unit[818.02]{keV} & \unit[8.59]{\%} &  \unit[0.24]{\%} & 16.3 &\baseT{6.63}{18} \\
%\midrule
%\nuc{Pd}{110} $2^+_3$ \unit[1783.48]{keV} \\
%				\unit[1783.48]{keV} & \unit[21.57]{\%} &  \unit[0.88]{\%} & 6.2 & \baseT{6.45}{19} \\
%				\unit[1125.71]{keV} & \unit[78.43]{\%} &  \unit[2.48]{\%} & 12.0 &\baseT{9.41}{19} \\
%				\unit[657.76]{keV} & \unit[78.43]{\%} &  \unit[2.99]{\%} & 12.4 &\baseT{1.09}{20} \\
\mr
\mr
\multicolumn{ 4}{l}{\nuc{Pd}{102} $2^+_1$ \unit[475.1]{keV} decay mode:}\\
 				\unit[475.1]{keV} (\unit[100.0]{\%})  &   \unit[3.32]{\%}  & \unit[5.09]{\%} & \unit[2.75]{\%}\\
\mr
\multicolumn{ 4}{l}{\nuc{Pd}{102} $0^+_1$ \unit[943.7]{keV} decay mode:}\\
				 \unit[475.1]{keV} (\unit[100.0]{\%}) &   \unit[2.72]{\%} & \unit[4.31]{\%} & \unit[2.26]{\%} \\
				 \unit[468.6]{keV} (\unit[100.0]{\%}) &   \unit[2.75]{\%} & \unit[4.32]{\%} & \unit[2.30]{\%} \\
\mr
\multicolumn{ 4}{l}{\nuc{Pd}{102} $2^+_2$ \unit[1103.05]{keV} decay mode:}\\
				\unit[475.1]{keV}  (\unit[62.9]{\%})   &  \unit[1.76]{\%} & \unit[2.67]{\%} 	& \unit[1.48]{\%} \\
				\unit[627.9]{keV}  (\unit[62.9]{\%})   &  \unit[1.63]{\%} & \unit[2.54]{\%} 	& \unit[1.39]{\%} \\
				\unit[1103.1]{keV} (\unit[37.1]{\%})   &  \unit[1.04]{\%} & \unit[1.60]{\%} 	& \unit[0.83]{\%} \\
\br
\end{tabular}\\ 
\raggedright\footnotesize $^a$ typographical error in Ref. \cite{Lehnert:2013ch}\\
\end{table}

%
%\end{thebibliography}

\end{document}